\documentclass[twocolumn,amsmath,amssymb,aps,superscriptaddress,pre,10pt]{revtex4-1}
\usepackage{changes}
\usepackage{graphicx}
\usepackage{dcolumn}
\usepackage{bm}
\usepackage{color}
\usepackage{relsize}

\newif\ifgraph

\graphtrue             %

\begin{document}
\title{
Chemotaxis of Artificial Microswimmers in Active Density Waves}

\author{Alexander Geiseler}
\email[Corresponding author:\\]{alexander.geiseler@physik.uni-augsburg.de}
\affiliation{Institut f\"ur Physik, University of Augsburg, D-86159, Germany}

\author{Peter H\"anggi}
\affiliation{Institut f\"ur Physik, University of Augsburg, D-86159, Germany}
\affiliation{Nanosystems Initiative Munich, Schellingstra{\ss}e 4, D-80799 M\"unchen, Germany}
\affiliation{Department of Physics, National University of Singapore, 117551 Singapore, Republic of Singapore}

\author{Fabio Marchesoni}
\affiliation{Center for Phononics and Thermal Energy Science,
 School of Physics Science and Engineering, Tongji University, Shanghai 200092,
 People's Republic of China}
\affiliation{Dipartimento di Fisica,
Universit\`{a} di Camerino, I-62032 Camerino, Italy}

\author{Colm Mulhern}
\affiliation{Institut f\"ur Physik, University of Augsburg, D-86159, Germany}

\author{Sergey Savel'ev}
\affiliation{Department of Physics, Loughborough University,
Loughborough, LE11 3TU, United Kingdom}

\date{\today}

\begin{abstract}
Living microorganisms are capable of a tactic response to external
stimuli by swimming towards or away from the stimulus source; they do
so by adapting their tactic signal transduction pathways to the
environment. Their self-motility thus allows them to swim against a
traveling tactic wave, whereas a simple fore-rear asymmetry argument
would suggest the opposite.  Their biomimetic counterpart, the artificial microswimmers, also propel themselves by harvesting kinetic energy from an active medium, but, in contrast, lack the adaptive capacity. Here we investigate the transport of artificial swimmers subject to traveling active waves and show, by means of analytical and numerical methods, that self-propelled particles can actually diffuse in either direction with respect to the wave, depending on its speed and waveform. Moreover, chiral
swimmers, which move along spiraling trajectories, may diffuse
preferably in a direction perpendicular to the active wave. Such a
variety of tactic responses is explained by the modulation of the
swimmer's diffusion inside traveling active pulses.
\end{abstract}
\maketitle

\section{Introduction} \label{intro}

Taxis is the biased movement of bacteria, somatic cells, or
multicellular organisms in response to external stimuli such as
light, electric currents, gravity and chemicals. Taxes are classified
based on the type of activating stimulus and on whether the organism's
resultant drift is oriented towards (positive) or away from
(negative) the stimulus source \cite{Murray}. Tactic cell migration
plays a crucial role in biological pattern formation and the
organization of complex biological organisms. For instance, bacteria
find food (e.g., glucose) or flee from poison (e.g., phenol) by
swimming respectively up (positive chemotaxis) or down (negative
chemotaxis) the concentration gradient of the sensed chemical
\cite{Berg,Gompper}.

A biomimetic counterpart of cellular motility is the ability of
specially designed synthetic microparticles to propel themselves by
harvesting kinetic energy from an active environment
\cite{Schweitzer,Reviews}. Self-propulsion is fueled by stationary
non-equilibrium processes, like directional ``power-strokes" from
catalytic chemical reactions or self-phoresis by short-scale
(electric, thermal, or chemical) gradients, produced by the particle
itself, in virtue of some built-in functional asymmetry
\cite{Schweitzer,Reviews,Sen_rev}. Similarly to bacteria, artificial
microswimmers are also known to diffuse up or down long-scale
monotonic gradients of the active medium \cite{SenPRL,Sano,Baraban,ChenChen}. However,
bacteria regulate their response to an external stimulus by adapting
their (complex) tactic signal transduction pathways \cite{Armitage},
which thus operate like sensor-actuator loops. In contrast, due to
their often sub-micron size and lack of an internal structure,
synthetic swimmers are unable to process the tactic signal
\cite{rmpHM_2009,chemphyschem}, i.e. their response being strictly
local.

\begin{figure*}[tp]
\centering
\includegraphics[width=0.72\textwidth]{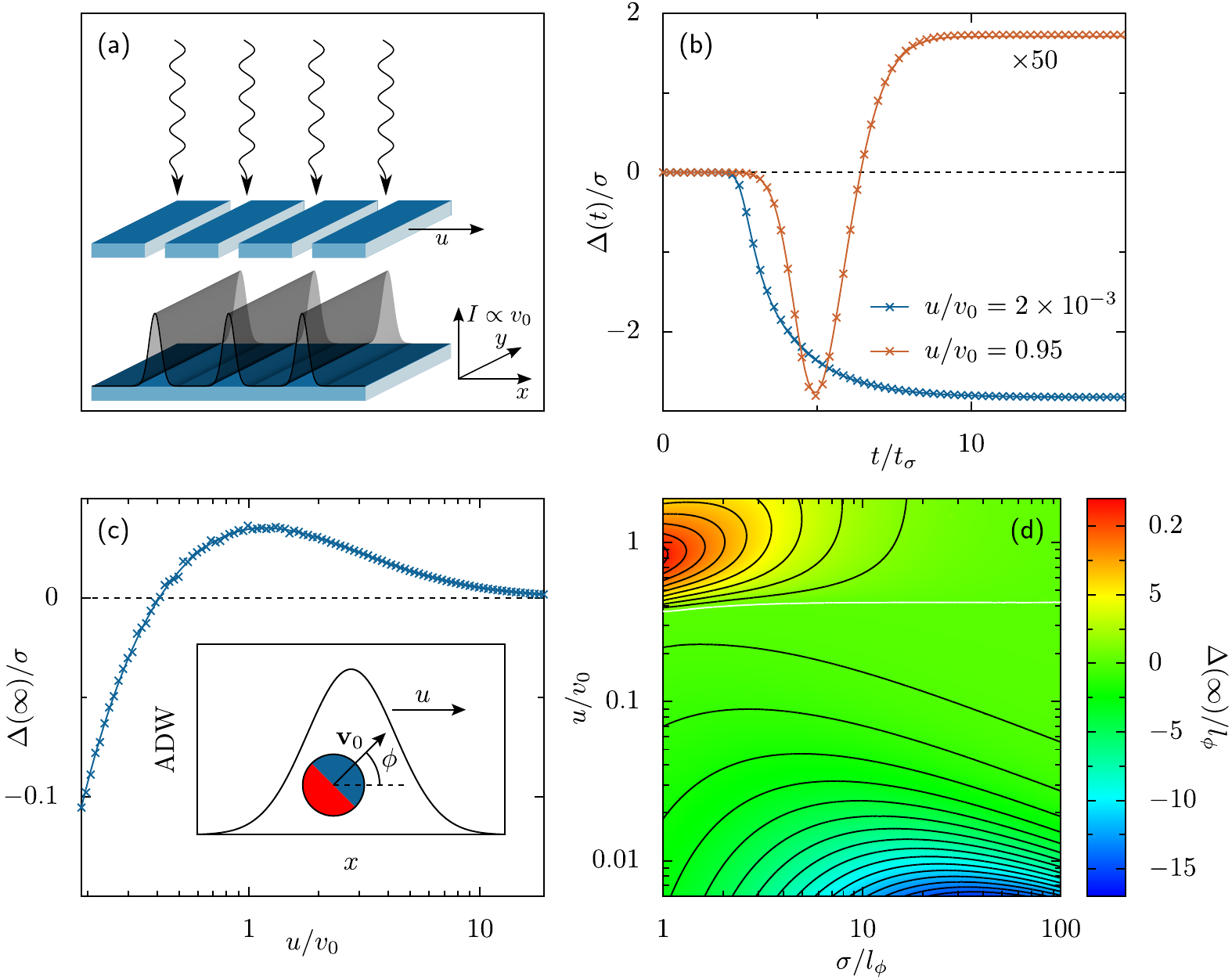}
\caption{(a) Ideal experimental setup of a thermophoretic swimmer
diffusing on a planar substrate irradiated by a laser beam
\cite{Sano,Bechinger}. By pulling at constant
speed a slit-screen sliding between the laser and the substrate, it is
possible to modulate the laser intensity, $I$, hitting the particle,
thus realizing an effective ADW. (b),(c) Chemotactic shift induced by a Gaussian active pulse, $v(x)
=v_0 \exp(-x^2/2\sigma^2)$, traveling to the right with constant
speed $u$ (see \ inset of (c), where a simple sketch of the model is
depicted). The swimmer's self-propulsion parameters, $v_0=53\,{\rm\mu
m/s}$ and $D_\phi=165\,{\rm s^{-1}}$, were chosen to mimic the
experimental setup of Ref.\ \cite{Bechinger}, and the pulse width,
$\sigma = 1\,{\rm\mu m}$, was set three times the swimmer's
propulsion length $l_\phi=v_0/D_\phi$. Translational noise $D_0$ and chiral
torque $\Omega$ were set to zero to focus on the basic mechanism
responsible for the emergence of the spatial shift, $\Delta (t) =
\langle x(t)-x(0)\rangle$.  In (b) the particle displacement,
$\Delta(t)$, is plotted vs. $t$ in units of the pulse crossing time
$t_\sigma=\sigma/u$ for a slow and fast pulse. For the sake of
comparison, we remind that the time the particle takes to diffuse a
length $\sigma$ in the bulk is $\tau_\sigma=\sigma^2/2D_s$. In (c)
the final displacement $\Delta (\infty)$ is plotted as a function of
the pulse speed. The stochastic integration of the model Eqs.\
(\ref{LE1}) (crosses), are compared with the numerical solution the
corresponding Fokker-Planck Eq. (\ref{FPE}) (solid curves). The contour plot in (d) illustrates the dependence of $\Delta(\infty)$ on both pulse parameters $\sigma$ and $u$. The white curve separates the regions with positive and negative $\Delta (\infty)$. \label{F2}}
\end{figure*}

In most circumstances the tactic stimulus is modulated in space and
time in the form of single or entrained active pulses that sweep
through the suspended swimmer. Certain microorganisms are able to
locate the source of the pulses and move towards it, no matter what
the sign of their response to a monotonic active gradient. A study
case is the chemotactic aggregation of the amoebae of the cellular
slime mold (Dictyostelium discoideum) \cite{paradox1}. This process
is directed by periodic sequences (or waves) of {\it symmetric}
concentration pulses of a chemoattractant, irradiating from the
aggregation center outwards. As amoebae exhibit positive chemotaxis,
one would expect a cell movement towards the center in the wave-front
and away from it in the wave-back. As a consequence, amoebae would
spend more time in the back-wave than in the front-wave, thus
undergoing a net drift in the direction of the active wave
propagation. Most remarkably, the conclusion of this argument (first
proposed by Stokes in a more general fluidodynamics context
\cite{stokes1,stokes2}) does not change by reversing the sign of
amoeba chemotaxis: According to Stokes' wave fore-rear asymmetry
argument and contrary to experimental observations, swimmers with
definite chemotaxis (positive or negative, alike) would always surf
active density waves (ADW), thus moving away from the the wave source
\cite{stokes3}. A proposed resolution of this apparent paradox (termed ``chemotactic wave paradox" in Ref. \cite{paradox2}) for cellular microorganisms requires the standard model for cell chemotaxis to be modified so as to account for a {\it finite} adaptation time of the chemotactic pathways to temporally varying stimuli \cite{paradox2}.

This approach cannot be extended to the case of artificial swimmers
diffusing across a traveling ADW because these respond to the
instantaneous activation properties of the surrounding medium with no
temporal memory \cite{Kapral}. However, controlling the transport of synthetic sub-micron particles, self-propelling in a spatio-temporally
modulated active medium, is key to the success of nanorobotics in
technological applications, like environmental monitoring,
intelligent drug delivery, or even more challenging biomedical tasks
\cite{Sen_rev}. With this goal in mind, we numerically investigated
the diffusive dynamics of artificial microswimmers at low Reynolds
numbers, subjected to traveling symmetric active pulses of different
waveforms. We observed that swimmers with positive taxis in a
monotonic active gradient, actually drift towards or away from the
pulse source, depending on the pulse sequence. Moreover, chiral
swimmers, which due to some configurational asymmetry tend to move in
circles \cite{Lowen}, may also drift orthogonally to the incoming
active wave. The variability of swimmers' tactic response is proven
to result more from the spatio-temporal modulation of their active
motion within a traveling pulse, than the pulse fore-rear gradient
asymmetry.

The paper is organized as follows. In Sec.\ \ref{results} we introduce the model of the artificial microswimmer that is to be considered. Already in this section, without resorting to the finer technical details (given later), a qualitative presentation of the key results will be given. Here the tactic responses of the swimmer to single wave pulses and also to active density waves are discussed. A more quantitative discussion of these results is given in Sec.\ \ref{discussion}, where, in addition, related analytical results will be presented. Finally, in Sec.\ \ref{methods}, the details of the methods used to obtain the results will be explained, including the numerical schemes applied to integrate the Langevin and corresponding Fokker-Planck equations.

\section{Results}\label{results}

\subsection{Model}
Regardless of the details of the self-propulsion
mechanism, at low Reynolds numbers the diffusion of an artificial
microswimmer on a surface is conveniently modeled by a set of simple
Langevin equations (LE)
\begin{eqnarray}
\dot x&=&v(x,t)\cos \phi +\sqrt{D_0}\; \xi_x(t),\nonumber \\ \dot
y&=&v(x,t)\sin
\phi+\sqrt{D_0}\; \xi_y(t), \nonumber \\
\dot \phi&=&\Omega +\sqrt{D_\phi}\; \xi_\phi(t) \label{LE1}.
\end{eqnarray}
Three sources of fluctuations are explicitly incorporated in the
model: two translational of intensity $D_0$, and one orientational of
intensity $D_\phi$. All noises are Gaussian and stationary, with
zero-mean and autocorrelation functions $\langle \xi_i(t)
\xi_j(0)\rangle =2\delta_{ij}\delta(t)$, with $i,j=x,y,\phi$. As such
noises result from a combination of thermal fluctuations in the
suspension fluid and randomness of the propulsion mechanism, we treat
the intensities $D_0$ and $D_\phi$ as independent parameters.
Finally, depending on their geometry, 2D swimmers often experience an
additional torque, $\Omega$, which makes them rotate counter-clockwise
or clockwise. Positive and negative chiral effects impact the
transport properties of both biological and synthetic active swimmers
\cite{Lowen,tenHagen,LowenBechinger,EPJST}.

When the swimmer floats in a homogeneous active suspension, its
propulsion speed is a constant, $v(x)=v_0$. As detailed in Sec.\ \ref{methods},
it then undergoes an active Brownian motion with finite persistence
time, $\tau_\phi=1/D_\phi$, length, $l_\phi=v_0\tau_\phi$, and bulk
diffusion constant $\lim_{t\to \infty}\langle
[x(t)-x(0)]^2\rangle=D_0+D_s$, with $D_s=v_0^2/2D_\phi$.

To model the effects of an active pulse sweeping through the
suspension fluid, we assume that the propulsion speed, $v(x,t)$, is a
local function of the physio-chemical properties of the medium at the
swimmer's position. For pulses propagating from left to right along
the $x$ axis, this amounts to inserting in Eqs.\ (\ref{LE1}) an
appropriate function $v(x,t)=v(x-ut)$, where $u$ is the pulse's speed
and the waveform $v(x)$ is chosen so as to describe single or
entrained traveling pulses \cite{Schweitzer} with amplitude $v_0$.
Like in Ref. \cite{paradox2}, here we restrict our analysis
to spatially symmetric waveforms, $v(x)=v(-x)$, in order to avoid
additional ratchet effects \cite{stokes3,rmpHM_2009}. The exact
mechanism underlying the tactic stimulus modeled by the spatio-temporal
function $v(x,t)$ does not, in principle, need to be specified; it can, for instance, be of {\mbox{chemo-,} \mbox{diffusio-,} electro- or
thermophoretic nature.

A proof of concept of the tactic effects predicted in this work is sketched in Fig.\ \ref{F2}(a). In this ideal experiment a thermophoretic swimmer, floating on a planar substrate, is irradiated by a defocused laser beam \cite{Sano}. A traveling train of laser intensity pulses is created by inserting a slit screen between the light source and the substrate and sliding it at a constant speed. Since the self-propulsive speed of a thermophoretic swimmer is approximately proportional to the laser intensity \cite{Buttinoni}, one can thus tailor at will the spatio-temporal modulation of the velocity field, $v(x,t)$.

For the sake of concreteness, in the following we adopt the term positive or negative chemotaxis to mean the drift of an active swimmer parallel or
anti-parallel to the direction of a generic incoming active pulse or
wave, regardless of its nature. Moreover, we remind the reader that, like in the experimental setup of Fig.\ \ref{F2}(a),
the propulsion speed of most artificial chemotactic swimmers grows
linearly with the concentration of the chemoactivants in the active
suspension, whereas, as assumed in Eqs.\ (\ref{LE1}), their angular
diffusion stays almost the same \cite{Sen_rev}. Such swimmers, when placed in a static velocity field, that is for
$v(x,t)=v(x)$, are known to diffuse up the velocity gradient
\cite{SenPRL,Nori}. Other authors \cite{Sano,Golestanian} have
detected the dependence of the rotational diffusion on the active
density wave to be generally weaker than of the propulsion speed.
Extending our analysis to an $x$-dependent $D_\phi$ would not alter
the general conclusions of the present work, except for some more
laborious technical details \cite{Schnitzer}.

Finally, we stress that in the ideal experiment of Fig.\ \ref{F2}(a) hydrodynamic effects can be largely suppressed, at least in the absence of activant gradients. Indeed \cite{Gompper}, (i) swimmers freely diffuse in the bulk away from the container's walls; (ii) their density can be lowered so as to avoid clustering \cite{Navarro}}; and (iii) they can be fabricated rounded in shape and so small in size (i.e., pointlike) to minimize hydrodynamic backflow effects. On the contrary, the activant gradients considered in the present work surely cause additional hydrodynamic effects in the form of a polarizing torque that tends to align the particle's velocity parallel or anti-parallel to the gradient, depending on the swimmer's surface properties \cite{Wurger}. Our simulations show that chemotaxis is enhanced for swimmers aligned against the activant gradient and suppressed in the opposite case, as also reported in Ref. \cite{Sano}. However, we numerically checked that the chemotactic effect persists even in the latter case, solely its magnitude slightly diminishes. To keep our discussion as simple as possible, polarization effects will be neglected here and fully investigated in a follow-up technical report.

\begin{figure*}
\includegraphics[width=0.72\textwidth]{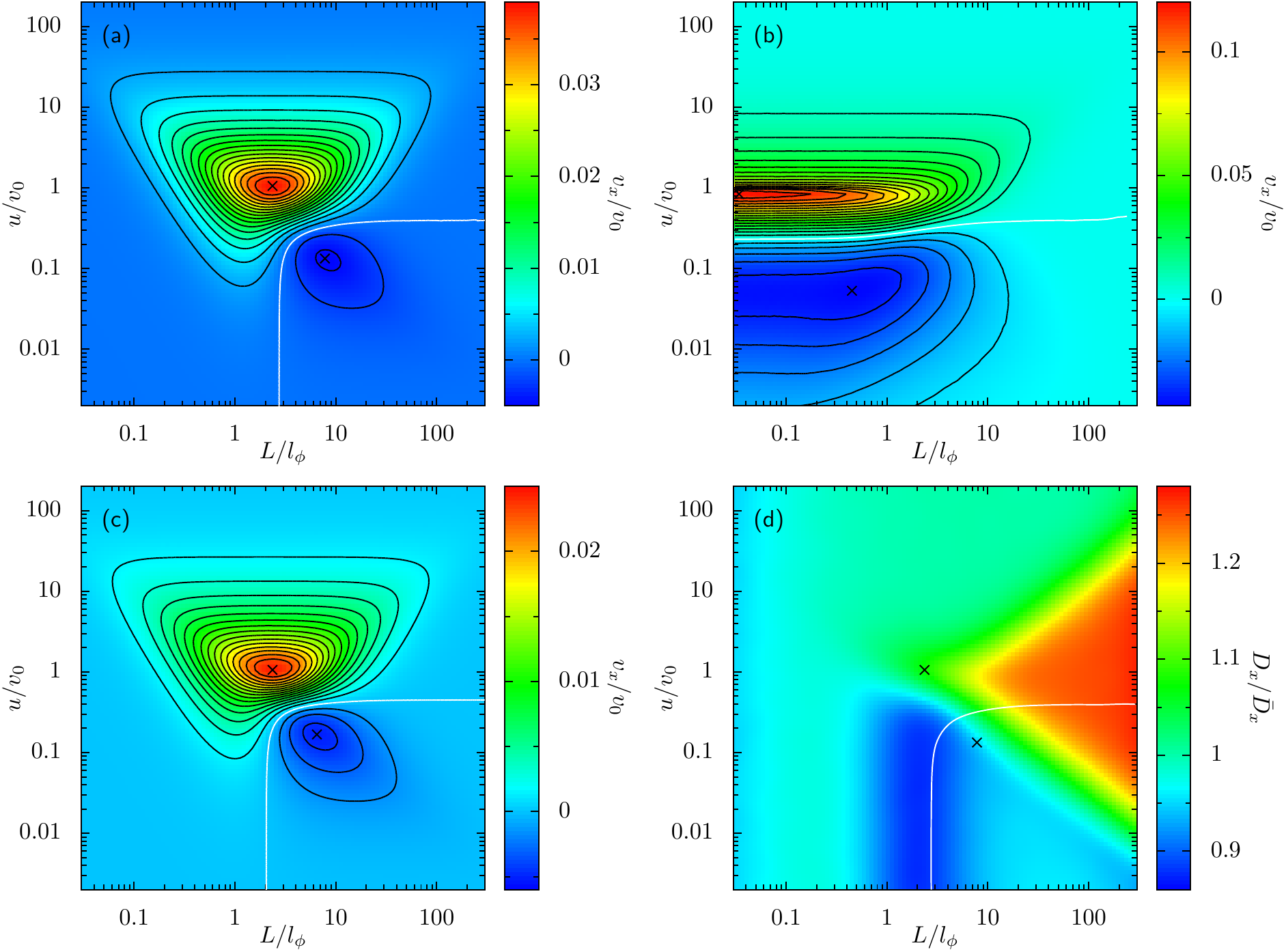}
 \caption{Chemotaxis of an achiral swimmer with
$\Omega=0$ in a sinusoidal traveling ADW. Unless differently stated,
all model parameters were chosen consistently with reported
experimental values \cite{Bechinger}: $v_0=53\,{\rm\mu m/s}$,
$D_\phi=165\,{\rm s^{-1}}$, and $D_0=2.2\,{\rm\mu m^2/s}$. The ADW
waveform, $v(x)=w_0+(v_0-w_0)\sin^2(\pi x/L)$, has fixed maxima,
$v_0$, and tunable minima $w_0$, with $w_0 < v_0$. All results are
from numerical integration of the model Eqs.\ (\ref{LE1}) or the
corresponding FPE (\ref{FPE}) (see Sec.\ \ref{methods}). Lengths and velocities
are given in units of $l_\phi$ and $v_0$, respectively. (a)-(c)
Contour plots of the longitudinal drift velocity, $v_x=\lim_{t\to
\infty} \langle x(t) -x(0)\rangle/t$ in the plane $(L,u)$; $w_0=0$
and all other parameters are the same except for $D_0=0$ in (b) and
$w_0=0.2$ in (c). The maxima of positive, $v_{x,\rm max}$, and
negative, $v_{x,\rm min}$, chemotaxis are marked by black crosses;
the separatrix curves, $u_s(L)$, delimiting the regions of positive
and negative chemotaxis are drawn in white;  (d) Contour plot of the
diffusion constant $D_x= \lim_{t\to \infty} \left[\langle x^2\rangle
-\langle x \rangle^2\right]/2t$ for the model parameters of panel
(a). The scaling factor, $\bar D_x=D_0+v_0^2/8D_\phi$, is the
swimmer's diffusion constant in the average velocity field $\langle
v(x)\rangle=v_0/2$. For animations of the swimmer's diffusion under
diverse dynamical conditions see Supplementary Movies 3-6 \cite{SI}. \label{F3}}
\end{figure*}

We now qualitatively discuss the numerical results of Figs.
\ref{F2}-\ref{F4}, that were obtained by numerically integrating the
LEs (\ref{LE1}) and the corresponding Fokker-Planck equation (FPE)
(see Sec.\ \ref{methods}). A more technical analysis of these results will be
presented in Sec.\ \ref{discussion}.

\subsection{Chemotaxis by single active pulses}
In Figs.\ \ref{F2}(b)-(d) and Supplementary Movies 1 and 2 \cite{SI} we
illustrate the effects of an active Gaussian pulse of amplitude $v_0$
traveling from left to right across a swimmer at rest in the absence
of translational fluctuations, $D_0=0$. The swimmer undergoes a net
longitudinal shift, $\Delta(\infty)=\lim_{t\to\infty}\langle
x(t)-x(0)\rangle$: most
notably, $\Delta(\infty)$ is markedly negative for slow pulses, $u
\ll v_0$, and positive for fast pulses, $u \geq v_0$, see Figs.\ \ref{F2}(c),(d). We explain the
existence of opposite chemotactic regimes by noticing that a 
swimmer with $D_0=0$ diffuses only across the width of the incoming active pulse. Suspended inside a slow travelling pulse with $u \ll v_0$, the particle quickly diffuses either against the front or the rear of the pulse. Upon reaching either pulse's edge, its diffusivity gets suppressed, that is its self-propulsive velocity, $v(x)$, grows smaller than the pulse speed, $u$. The ensuing behaviour at the two sides of the pulse is different.
After hitting the r.h.s., the particle is caught up again by the advancing pulse and resumes diffusing, whereas upon hitting the l.h.s., it is left behind and comes to rest. The two pulse's edges behave, respectively, like travelling reflecting and adsorbing walls. As a result we expect, on average, a net shift of the particle to the left. In the opposite regime of a fast travelling pulse, $u \gg v_0$, the particle comes almost
immediately to rest when hitting the left pulse wall, while it can
travel a much longer distance to the right without bouncing against
the right pulse wall. In the optimal case, $u \simeq v_0$, such a
distance is of the order of the persistence length, $l_\phi$.
Accordingly, for a fixed pulse width, $\Delta(\infty)$ attains a positive maximum around $u
\simeq v_0$, and vanishes monotonically in the limit $u/v_0 \to
\infty$.

Consistent with our interpretation, in Figs.\ \ref{F2}(c)-(d) the
transition from negative to positive chemotaxis occurs when the time
the particle takes to diffuse a length of the order of $l_\phi$, grows longer than the corresponding pulse crossing time, namely for $u/v_0
\simeq 1/2$. Of course, under realistic experimental conditions, the apparent divergence of the shift, $\Delta(\infty) \to -\infty$, at vanishingly slow pulse speeds, Fig.\ \ref{F2}(c), would be offset by the inevitable translational fluctuations with $D_0>0$, neglected in the simulations of Fig. \ref{F2}. Moreover, for narrow pulses, $\sigma\lesssim l_\phi$, the swimmer cannot diffuse much inside the pulse, but crosses it  \emph{ballistically}. Consequently,  as illustrated in Fig.\ \ref{F2}(d), its positive drift is more pronounced than in the case $\sigma\gg l_\phi$. Vice versa,  for large pulse widths the particle's dynamics is dominated by active diffusion and its displacement grows increasingly negative at small $u$. Of course, for $\sigma\to\infty$ the activant gradient becomes negligible and $\Delta(\infty)$ vanishes. Thus, in the regime of negative $\Delta(\infty)$, for any chosen $u$ there exists an optimal pulse width where the negative particle displacement is ma!
 ximum.

\subsection{Chemotaxis by active density waves}
More generally, active pulses are generated in random or periodic sequences. For the sake of simplicity,
we consider here the case of periodic ADWs with waveform
$$v(x)=w_0+(v_0-w_0)\sin^2(\pi x/L).$$
Like in most experimental
setups, we assume that the parameters that regulate the swimmer's
dynamics, $D_0, v_0$, $\tau_\phi$ and $l_\phi$, are fixed, whereas
the ADW parameters, $u, L$ and $w_0/v_0$, can be tuned at the
experimenter's convenience. We checked that the swimmer's chemotactic
response is not appreciably modified by varying the sequence or
waveform of the active pulses.

\begin{figure}[tp]
\includegraphics[width=0.86\columnwidth]{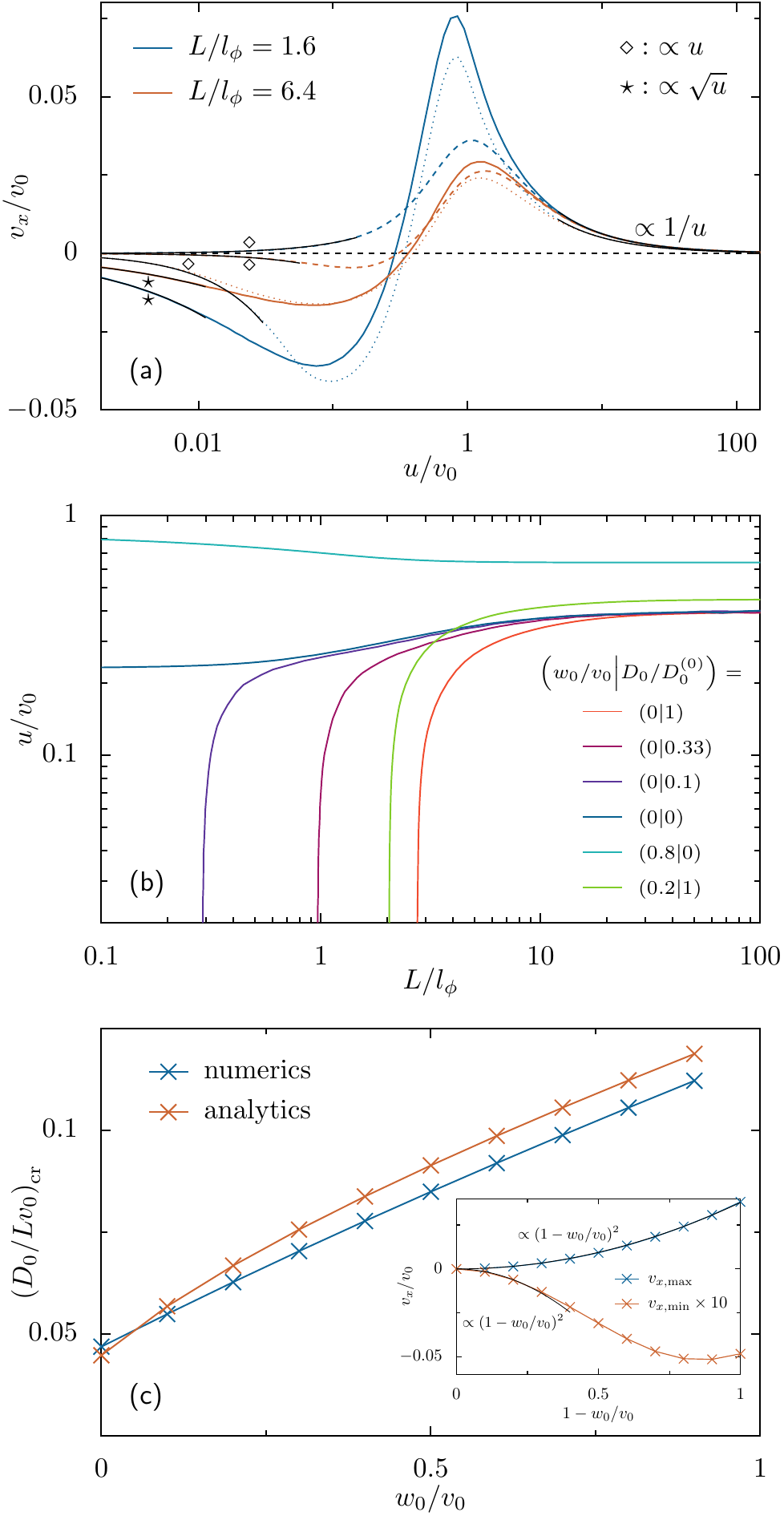}
\caption{Chemotaxis of an achiral swimmer in a sinusoidal traveling
ADW. Unless differently stated, all parameters were chosen as in
Fig.\ \ref{F3}(a). Lengths and velocities are given in units of
$l_\phi$ and $v_0$, respectively. (a) $v_x$ vs. $u$ in the ballistic
regime, $L/l_\phi=1.6$,  and diffusive regime, $L/l_\phi=6.4$, for
finite $D_0$ and $w_0=0$ (dashed curves), $D_0=0$ and $w_0=0.1$
(dotted curves) and $D_0=0$ and $w_0=0$ (solid curves). All curves
decay asymptotically with power laws, as indicated; (b) separatrix,
$u_s$ vs. $L$ for different values of $D_0$ and $w_0$, where
$D_0^{(0)}$ denotes the standard value of $2.2\,\mu\mathrm{m}^2/s$
used in the previous figures. The vertical branch of $u_s$ shifts to
lower $L$ proportional to the noise level, $D_0$, and inversely
proportional to the ADW baseline, $w_0$; (c) $(D_0/Lv_0)_{\rm cr}$
vs. $w_0$ from the numerical integration of the exact Eq.\
(\ref{FPE}) (blue crosses), and the approximated FPE in the ballistic
regime (red crosses), see Eq.\ (\ref{vxballistic}). In the inset the
maxima and minima of $v_x$, $v_{x,\rm max}$ and $v_{x,\rm min}$, are
shown to increase faster than linearly with the amplitude of the ADW,
$v_0-w_0$. Note, however, that the modulus of $v_{x,\rm min}$
increases with $w_0$ for $w_0 \ll v_0$, as also shown in (c).
\label{F5}}
\end{figure}

In the stationary regime, chemotaxis of achiral swimmers with
$\Omega=0$ is characterized in terms of the drift velocity,
$v_x=\langle \dot x \rangle$. This is plotted in Fig.\ \ref{F3} for
different values of the wave parameters. The regions of the plane
$(L,u)$ exhibiting positive or negative chemotaxis are delimited by
{\it separatrix} curves, which depend also on $D_0$ and $w_0$ [Figs.
\ref{F3}(a)-(c) and \ref{F5}(b)]. Like in the case of single active
pulses, negative chemotaxis is induced by slow ADWs only, with
$u<v_0$ (Supplementary Movies 3-6 \cite{SI}). Indeed, for fast ADWs, the distance a swimmer can travel
without crossing a minimum of $v(x)$ is longer to the right than to
the left. Hence, one can expect that $v_x>0$. Moreover, the distance
the swimmer can surf a wave with $u \simeq v_0$ is limited solely by
its persistence time, $\tau_\phi$, so that here positive chemotaxis is
the most pronounced. Vice versa, for $u\ll v_0$, the swimmer crosses
the ADW troughs, with reduced self-propulsion speed, both left and
right. As the ADW propagates to the right, the time the swimmer takes
to cross a trough to its left is shorter than to its right, hence
$v_x<0$ (Supplementary Movies 3 and 4 \cite{SI}). This argument certainly holds
for $D_0=0$, where---provided $w_0=0$---the particle can never cross
a trough to the right, see Fig.\ \ref{F3}(b) and Supplementary Movies 4 and 6 \cite{SI}.

The translational fluctuations, $D_0>0$, help the swimmer diffuse across the
ADW troughs, suppressing the velocity rectification mechanism
described above. Thus, with increasing $D_0$, the negative
rectification effect becomes smaller compared to the positive surfing
effect, until eventually $v_x$ changes sign. As the same effect can
not only be achieved by raising the noise strength $D_0$, but also
by lowering the pulse periodicity $L$ (translational
noise can easily ``kick'' particles out a narrow pulse), the separatrix bends downward, almost vertically, at a critical value of $L$ -- see Fig.\
\ref{F3}(a). The positive chemotaxis observed in the bottom-left
quadrant of Fig.\ \ref{F3}(a) is an unavoidable effect of the translational
fluctuations $D_0 \neq 0$.

The dependence of the vertical branch of the separatrix on the model
parameters is illustrated by the curves of Figs.\ \ref{F5}(b)-(c). We
observe that its position along the horizontal axis is (i) shifted to
the right proportional to $D_0$; (ii) shifted to the left
proportional to $w_0$; and (iii) independent of $l_\phi$ (not shown).
This behavior points to the existence of a critical value of
$D_0/Lv_0$, $(D_0/Lv_0)_{\rm cr}$, above which negative chemotaxis is
suppressed. Moreover, Fig.\ \ref{F5}(c) clearly shows that
$(D_0/Lv_0)_{\rm cr}$ grows linearly with $w_0$, independently of
$l_\phi$. This implies that the vertical branch of the separatrix can
be shifted to lower $L$ either by lowering $D_0$ or increasing $w_0$.
However, these two options for enlarging the negative chemotaxis
region of the $(L,u)$ plane, have opposite impact on the modulus of
$v_x$ -- chemotaxis is enhanced by lowering $D_0$ and suppressed by
raising $w_0$ at constant $v_0$ [Fig.\ \ref{F5}(c), inset].

The horizontal branch of the separatrix also depends on $w_0$, but is
insensitive to the noise intensities $D_0$ and $D_\phi$. In the limit
$D_0\to 0$, the separatrix is a smooth function of $L$, $u_s(L)$,
with limits $u_s(0)$ and $u_s(\infty)$ of the same order of
magnitude, both limits being functions of $w_0$ and smaller than
$v_0$. In view of these results we conclude that negative chemotaxis
is a robust property of the system, since it sets in under the most
affordable experimental conditions of ADWs traveling with low speed,
$u\ll v_0$, and long wavelength, $L\gg l_\phi$.

\subsection{Transverse chemotaxis of chiral swimmers}
\begin{figure*}[tp]
\includegraphics[width=0.903\textwidth]{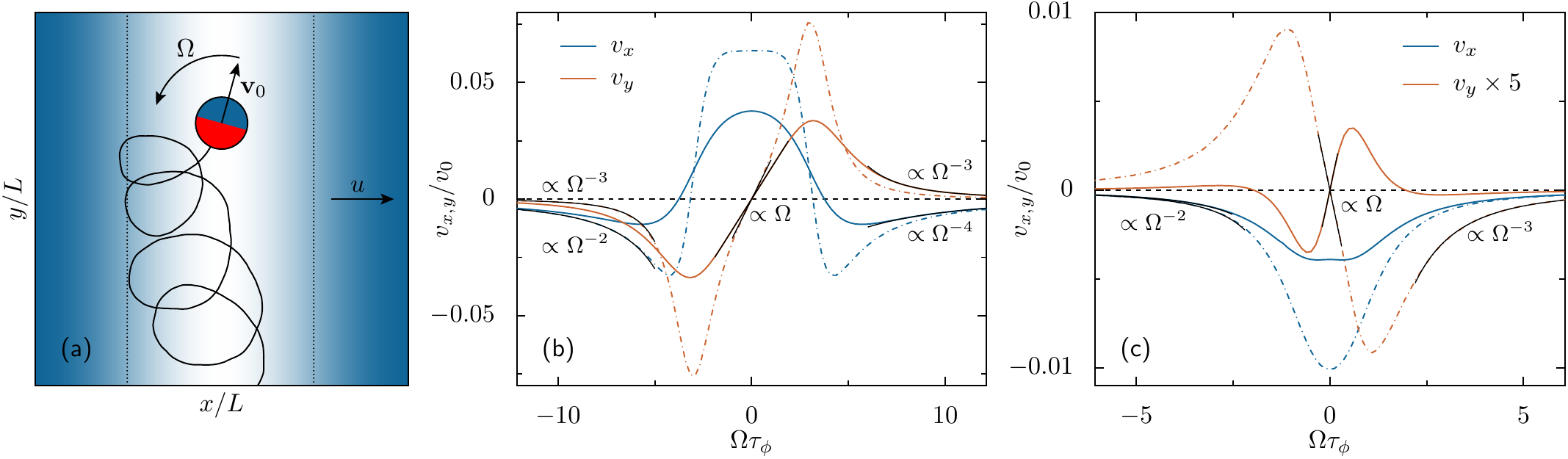}
 \caption{Chemotaxis of a chiral swimmer with
$\Omega\neq 0$ in a sinusoidal traveling ADW. Unless differently
stated, all model parameters were chosen as in Fig.\ \ref{F3}(a). All
results are from numerical integration of the model Eqs.\
(\ref{LE1}) or the corresponding FPE (\ref{FPE}). The drift velocities and chiral frequency are given in units of $v_0$ and $D_\phi=\tau_\phi^{-1}$, respectively. For animations see Supplementary Movies 7-10 \cite{SI}. In (a) is
a sketch of the upward spiraling trajectory of a swimmer bouncing
against the left trough of an ADW, where the dotted black line
indicates the half-width. The longitudinal and transverse drift
velocities, $v_x=\lim_{t\to \infty} \langle x(t) -x(0)\rangle/t$ and
$v_y=\lim_{t\to \infty} \langle y(t) -y(0)\rangle/t$ are plotted vs.
$\Omega \tau_\phi$ for (b) $L/l_\phi=2$, $u/v_0=1$; and (c)
$L/l_\phi=7$, $u/v_0=0.2$ (solid curves).  The corresponding curves
for $D_0=0$ are also drawn for the sake of comparison (dashed
curves). The exponents of fitted power-laws coincide with those
predicted in Sec.\ \ref{discussion}. \label{F4}}
\end{figure*}

An intrinsic rotational torque of the swimmer can be either the accidental result
of fabrication defects or the desired effect obtained, e.g., by
bending an active nanorod \cite{Takagi}. In any case, chirality
strongly impacts swimmers' chemotaxis. In Figs.\ \ref{F4}(b)-(c) the
longitudinal and transversal drift velocities, $v_x=\langle \dot
x\rangle$ and $v_y=\langle \dot y \rangle$, are plotted against the
torque frequency, $\Omega$, for conditions of, respectively, the
largest positive and negative chemotaxis at $\Omega=0$ [marked by
crosses in the contour plot of Fig.\ \ref{F3}(a)]. A few remarkable
properties are immediately apparent (see also Supplementary Movies 7-10 \cite{SI}): (i) Chirality induces a
transverse chemotactic drift, $v_y\neq 0$. Such an effect is the
strongest in the regime of positive longitudinal chemotaxis of
achiral particles; (ii) $v_x(\Omega)$ and $v_y(\Omega)$ are
respectively even and odd functions of $\Omega$, consistent with the
symmetry of the model Eqs.\ (\ref{LE1}) under the transformation $\phi
\to -\phi$; (iii) Chirality tends to suppress longitudinal
chemotaxis. This effect is best noticeable in Fig.\ \ref{F4}(b),
where for $\Omega \tau_\phi \sim \pi$
the longitudinal drift velocity, $v_x$, drops to zero, while $v_y$
develops a peak of height comparable with $v_x(0)$. Under these
conditions, the chemotactic effect of the incoming ADW is fully {\it
transverse} and with the same sign as $\Omega$. At even higher
frequencies, $v_x$ changes sign from positive to negative; (iv) In
the regime of negative chemotaxis of achiral particles and zero
translational noise [see\ Fig.\ \ref{F3}(b)], the curves $v_x(\Omega)$ and
$v_y(\Omega)$ have the same sign and no zeros for $\Omega>0$. On
raising $D_0$, without changing the sign of $v_x$, transverse chemotaxis is suppressed and its sign varies with increasing
$\Omega$.

The mechanism responsible for transverse chemotaxis is illustrated in
Fig.\ \ref{F4}(a). An active particle subject to a positive torque,
$\Omega >0$, bounces against the ADW troughs, thus tracing
spiraling circles of radius $r_d \sim v_0/\Omega$, which go up (down)
the right (left) hand side of the trough. For fast traveling ADWs
with $u > v_0$, bouncing trajectories only take place on the r.h.s.
of the wave troughs, so that $v_y>0$. Of course, transverse
chemotaxis is most pronounced when the bouncing process is
synchronized with the wave modulation, namely, when the bouncing
time, $\pi/\Omega$, is of the order of the wave period, $L/u$.
Moreover, such a mechanism grows more effective for swimmer
persistence times larger than the bouncing times, that is, for $\Omega
\tau_\phi \gtrsim \pi$. Both conditions are satisfied in Fig.
\ref{F4}(b), so that, in an appropriate $\Omega \tau_\phi$ range,
transverse chemotaxis supersedes longitudinal chemotaxis.

For slow ADWs, spiraling trajectories develop on both sides of the
wave minima. As noticed above, trough crossings from right to left
take a shorter time than from left to right; accordingly, upward
bouncing trajectory arcs have shorter span than the downward ones.
This observation explains why, in Fig.\ \ref{F4}(c) for $u \ll v_0$
and $D_0=0$, transverse chemotaxis and chirality have opposite signs.
Also, similarly to achiral chemotaxis, translational noise eases ADW trough
crossings in both directions, thus suppressing transverse chemotaxis
and eventually reversing its orientation. It is also suppressed by
reducing the ADW amplitude.

\section{Discussion} \label{discussion}

We next present a more quantitative analysis of our results based on
the approximation schemes detailed in Sec.\ \ref{methods}.

\subsection{Ballistic regime}
The curves of the drift velocity, $v_x$, versus the ADW speed, $u$, in Fig.\ \ref{F5}(a) exhibit a
characteristic resonant behavior for both positive and negative
chemotaxis. In the ballistic regime, i.e.\ for $L\ll l_\phi$, their
decay is satisfactorily described in the two-state approximation,
where, for sufficiently long persistence times (compared to the pulse
crossing time), the swimmer's dynamics is modeled as the
superposition of a non-fluctuating, self-propelling drift to the
right and to the left. In the absence of translational noise, the predicted
average drift velocity, Eq.\ (\ref{sol}), decays like: (i) $v_x/v_0
\propto v_0/u$ for $u\gg v_0$ (to hold for any value of $L/l_\phi$);
(ii) $v_x/v_0 \propto -u/v_0$ for $u \ll v_0$ and $w_0>0$; and (iii)
$v_x/v_0 \propto -\sqrt{u/v_0}$ for $u \ll v_0$ and $w_0=0$. These
predictions closely agree with the numerical fits of Fig.\
\ref{F5}(a). In addition, Eq.\ (\ref{sol}) suggests that the positive
chemotaxis maxima, $v_{x,\rm max}$, occur for $u \simeq v_0$, as
expected, and the negative minima, $v_{x,\rm max}$, for $u \lesssim
w_0$. The two-state model also provides a
close estimate of the horizontal branch of the separatrix, $u_s(L)$.
The limit $L/l_\phi \to 0$ for $D_0=0$, $u_s(0)$, can be approximated
analytically by setting $v_x=0$ in the second line of Eq.\
(\ref{sol}) and solving for $u$. Hence
$u_s(0)/v_0=[(w_0+v_0)/6v_0][1+\sqrt{1 +12 v_0w_0/(v_0+w_0)^2}]$, in
quantitative agreement with the numerical data of Fig.\ \ref{F5}(b).
The full curves $u_s(L)$ plotted there were obtained by locating the
zeros of the average current, Eq.\ (\ref{vdexact}), as a function of
$u$ and $L$.

As remarked in Sec.\ \ref{results}, translational fluctuations tend to suppress
chemotaxis at large, and negative chemotaxis in particular. In the
two-state model notation, this is a consequence of the noise-induced
``creeping effect" mentioned in Sec.\ \ref{methods}, an effect more conveniently
addressed in the FPE formalism. For instance, upon expanding the
ballistic approximation of the average current, Eq.
(\ref{vxballistic}), in powers of $u/v_0$, the chemotactic speed is
proven to grow proportionally to $u$ (and not $\sqrt{u}$) even at
$w_0=0$, with a sign that turns from negative to positive on
increasing $D_0$ [Fig.\ \ref{F5}(a)]. More interestingly, Eq.\
(\ref{vxballistic}) allowed us  to locate the vertical separatrix
branch of the contour plots of Figs.\ \ref{F3}(a)-(c), namely the
quantity $(D_0/Lv_0)_{\rm cr}$ plotted in Fig.\ \ref{F5}(c). For an
analytical estimate of the same quantity, we notice that translational
noise provides an additional diffusion mechanism that competes with
self-propulsion. The self-propulsion mechanism eventually prevails
when the time the particle takes to diffuse a half ADW
wavelength due to translational noise, $(L/2)^2/2D_0$, is shorter than the average time to cross
the same distance with an average speed $(v_0+w_0)/2$ in the
ballistic regime [see also Eq.\ (\ref{sol})], $2L/(v_0+w_0)$. This
occurs for $D_0/Lv_0 \gtrsim (D_0/Lv_0)_{\rm cr}$, with
$(D_0/Lv_0)_{\rm cr}\propto 1+w_0/v_0$, in agreement with the
numerical data in Fig.\ \ref{F5}(c). We remarked in Sec.\ \ref{results} that
$(D_0/Lv_0)_{\rm cr}$ is independent of $\tau_\phi$; therefore, this
analytical estimate holds in the ballistic and diffusive regimes,
alike.

Contrary to the Stokes drift \cite{stokes2}, $v_x$ at small $u$ grows
with exponent clearly smaller than $2$, irrespective of  $w_0$ and
$D_0$ (and, therefore, of its own sign): this means that the
chemotactic effect studied here is not governed by the fore-rear ADW
gradients so much as by the swimmer's diffusion across the traveling
ADW troughs.

\subsection{Diffusive regime}
In the diffusive regime, i.e.\ for $L\gg l_\phi$, the persistence time $\tau_\phi$ can be taken as vanishingly
small. Accordingly, the swimmer's diffusion in the wave direction is
closely described by the multiplicative LE in Eq.\ (\ref{LEdiff}).
Standard Stratonovitch calculus \cite{HT,Risken} yields the finite drift
term
\begin{equation} \label{Sdrift}
v_x= [(v_0-w_0)^2/4D_\phi]\langle (d/dx)\sin^2[\pi (x-ut)/L]\rangle,
\end{equation}
where $\langle \dots \rangle$ denotes a stationary average over the
stochastic trajectories of $x(t)$. This average was computed
explicitly by solving the corresponding FPE, Eq.\ (\ref{FPEdiff}):
For $u < v_0$ chemotaxis indeed turns out to be negative with $v_x
\propto -u$ for $D_0> 0$ and $u\ll v_0$, and $v_x \propto -\sqrt{u}$
for $D_0= 0$ and $u\ll v_0$, as in the fits of Fig.\ \ref{F5}(a).

On the other hand, upon increasing $u$ larger than the modulus of the
multiplicative term in Eq.\ (\ref{LEdiff}),  $u > |\langle v(x)\cos
\phi \rangle| \simeq (w_0+v_0)/2\sqrt{2}$, the dynamical effect of
the angular fluctuations is suppressed with respect to the dragging
action of the ADW, so that chemotaxis changes sign from negative to
positive. The r.h.s. of the above inequality approximates the
horizontal asymptote of $u_s(L)$ in the limit $L/l_\phi \to \infty$.
As displayed in Fig.\ \ref{F5}(b), $u_s(\infty)$ is a function of
$w_0$, only.

The role of the separatrix is further illustrated by the contour plot
of Fig.\ \ref{F3}(d), where we plotted the diffusion constant $D_x=
\lim_{t\to \infty} \left[\langle x^2\rangle -\langle x
\rangle^2\right]/2t$ for the model parameters of Fig.\ \ref{F3}(a).
The computed values of $D_x$ have been compared with the swimmer's
average diffusion constant, $\bar D_x$, defined in the figure
caption. One sees immediately that $D_x/\bar D_x \simeq 1$ almost
everywhere in the $(L,u)$ plane, except across the (hot) horizontal
and (cold) vertical branches of the separatrix, where $D_x/\bar D_x
>1$ and $D_x/\bar D_x < 1$, respectively. Indeed, the chemotaxis sign
inversion observed upon increasing the ADW speed, $u$, signals a
locked-running transition in the two-state model Eqs.\
(\ref{LE2state}), a mechanism known to produce excess diffusion
\cite{Costantini,Reimann,Reimann2}. On the other hand, the chemotaxis
inversion obtained by increasing the ADW wavelength is governed by
translational noise (it never occurs for $D_0=0$), which acts there as a
sort of lubricant, thus suppressing the net swimmer diffusivity. With a view to experimental demonstration, we remark that
for artificial swimmers, chemotaxis by traveling ADWs is no more
dispersive than regular transport in the bulk.

\subsection{Chiral chemotaxis}
The FPE formalism also allows a better characterization of chiral chemotaxis. To shed light on the
underlying irreducible 2D mechanism we considered the model Eqs.
(\ref{LE1}) in the opposite adiabatic limits $\Omega \tau_\phi \ll 1$
and $\Omega \tau_\phi \gg 1$, both in the ballistic and diffusive
regimes introduced above. A systematic perturbation approach (not
reported here) led us to conclude that $v_x=c_1(v_0-w_0)$ and
$v_y=s_1(v_0-w_0)$, where $c_1$ and $s_1$ are, respectively, even and
odd functions of $\Omega$. In particular, for $\Omega \tau_\phi \ll
1$ we obtained $c_1 \propto \Omega^0$ and $s_1 \propto \Omega$ for
any $D_0$; for $\Omega \tau_\phi \gg 1$, $c_1$ decays like
$\Omega^{-2}$ at $D_0=0$ and $\Omega^{-4}$ at $D_0\neq 0$, whereas
$s_1 \propto \Omega^{-3}$, independent of the translational noise strength $D_0$.

We conclude by underscoring that chemotaxis of artificial
microswimmers is a robust phenomenon that lends itself to accessible
laboratory demonstrations and promising applications to
nanotechnology and medical sciences. Our numerical results and
interpretation go beyond the earlier ``chemotactic wave paradox"
debate, insofar as the swimmer's tactic response results from its
ability of diffusing within a traveling active pulse and not just the
asymmetry of the fore-rear pulse gradients. Moreover, the direction
and magnitude of the tactic response are extremely sensitive to the
self-propulsion mechanism, which suggests the design of tactic
devices to control the production and transport of artificial
swimmers.

\section{Methods} \label{methods}

The model Eqs.\ (\ref{LE1}) describe the spatial diffusion of an
active over-damped Brownian particle subject to a traveling field of
force $v(x,t)$. Contrary to the previous literature on Brownian
Stokes' drift \cite{stokes2,stokes3}, here the particle's motion is
characterized by a finite persistence time, $\tau_\phi$, and related
length $l_\phi=v_0\tau_\phi$. A simple calculation \cite {Ghosh}
shows that the angular factors, $\cos \phi$ and $\sin \phi$, decay
exponentially, $\langle \cos \phi(t) \cos \phi(0) \rangle =\langle
\sin \phi(t) \sin \phi(0) \rangle=(1/2)\exp(-D_\phi |t|)$, hence
$\tau_\phi=1/D_\phi$. Persistence (or memory effects) become
appreciable when the Brownian particle is confined to geometries with
characteristic size smaller than $l_\phi$. In our model, the standard
distinction between a {\it ballistic} regime with $L\ll l_\phi$, and
a diffusive regime with $L\gg l_\phi$, must be revised to account for
the finite wave period $L/u$, as discussed below.

\subsection{The Langevin equations}
The stochastic differential Eqs.\ (\ref{LE1}) were numerically integrated by means of a standard
Euler-Maruyama scheme \cite{Kloeden}. The stochastic averages were
taken over an ensemble of trajectories with random initial swimmer
orientation, $\phi(0)\in [0,2\pi]$.

{\it The two-state model.} In the ballistic regime, the analytic
treatment of Eqs.\ (\ref{LE1}) with $\Omega=0$ is greatly simplified
by assuming that an achiral swimmer moves in the wave's direction
with equal probability to the right, $\phi=0$, or to the left,
$\phi=\pi$, i.e. with velocity $\pm v(x,t)$, thus totally ignoring
its transverse motion along $y$. The ensuing swimmer dynamics is
modeled through two independent LEs
\begin{equation}\label{LE2state}
\dot x= \pm w_0 \pm (v_0-w_0)\sin^2[\pi(x-ut)/L]
+\sqrt{D_0}~\xi_x(t),
\end{equation}
obtained by setting $\cos \phi = \pm1$ in Eq.\ ({\ref{LE1}}). By
introducing the auxiliary variable $x'=x-ut$, Eqs.\ (\ref{LE2state})
can rewritten in the form of two LEs for for a tilted potential with
different tilting,
{\small
\begin{equation}\label{LE2wash}
\dot x'= -u \pm [(v_0+w_0) - (v_0-w_0)\cos(2\pi x'/L)]/2
+\sqrt{D_0}~\xi_x(t).
\end{equation}
}%
The time a noiseless swimmer takes to cross a wavelength $L$ can be
calculated analytically from Eqs.\ (\ref{LE2wash}) with $D_0=0$. In
the steady-state with $\phi=0$ (oriented to the right), we obtained
$\dot x'>0$ for $0\leq u \leq w_0$, $\dot x'=0$ for $w_0 \leq u \leq
v_0$, and $\dot x'<0$ for $u\geq v_0$; in the steady-state with
$\phi=\pi$ (oriented to the left), $\dot x'<0$ for all $u$. The
coordinate $x'$ is thus locked for $\phi=0$ and $w_0 \leq u \leq
v_0$, and running under all remaining conditions, with
$|v_{x'}^{(\pm)}| = L/t_{\pm}$ for $\phi=0$ and $\phi=\pi$,
respectively, where $t_{\pm}=L/\sqrt{(v_0\mp u)(w_0 \mp u)}$.
Accordingly, on transforming back to the coordinate $x$, we obtain
two distinct solutions for $\langle \dot x \rangle$, $v_x^{(+)}$ and
$v_x^{(-)}$. Finally, averaging over the two $\phi$ states
corresponds to taking the arithmetic mean
$v_x=[v_x^{(+)}+v_x^{(-)}]/2$, that is
\textcolor{black}
{\small
\begin{equation}\label{sol}
v_x=\left\{\begin{array}{l}
u+\frac{1}{2}\left[\sqrt{(v_0-u)(w_0-u)}-\sqrt{(v_0+u)(w_0+u)}\right]\\
0\leq u\leq w_0\\
u-\frac{1}{2}\sqrt{(v_0+u)(w_0+u)}\quad\quad w_0\leq u\leq v_0\\
u-\frac{1}{2}\left[\sqrt{(u-v_0)(u-w_0)}+\sqrt{(u+v_0)(u+w_0)}\right]\\
u\geq v_0
\end{array}\right.
\end{equation}
}%
As shown in Sec.\ \ref{discussion},
despite the rather rough assumptions detailed above, this result
proves to be an effective interpretation tool.

Such a ballistic scheme holds in a strict sense under the condition
that $\tau_\phi > {\rm max}\{t_+, t_- \}$, namely for $u/v_0 >
[\sqrt{(2L/l_\phi)^2+(1-w_0/v_0)^2} +(1+w_0/v_0)]/2$. This implies
that  Eq.\ (\ref{sol}) applies for $u>v_0$ when $L/l_\phi \ll 1$, and
for $u/v_0 > L/l_\phi$ when $L/l_\phi \gg 1$. Moreover, since in the
state $\phi=0$ with $w_0 < u < v_0$ the particle  is locked, i.e.
$t_+=\infty$, the solution for $v_x$ holds approximately under the
weaker condition $\tau_\phi> t_-$. This is why in Sec.\ \ref{discussion} we
extended our two-state model interpretation to $u \ll v_0$ and $L \ll
l_\phi$.

To solve the two-state model of Eqs.\ (\ref{LE2state}) in the presence
of translational noise we had recourse to the FPE formalism, as discussed
below. We anticipate that for $D_0>0$ one obtains $v_{x'}^{(+)}>0$
also in the range $w_0 \leq u \leq v_0$ \cite{Risken}: translational
noise smooths the locked-running transition, thus acting as a
dynamical lubricant (creeping effect).

{\it 1D reduced model.} Brownian motion can be treated as purely
diffusive when its persistence length is much shorter than all other
length scales of the system. For the swimmer of Eq.\ (\ref{LE1}) this
condition amounts to requiring that the persistence time,
$\tau_\phi$, is shorter than both crossing times $t_+$ and $t_-$,
namely for $u/v_0 \leq [\sqrt{(2L/l_\phi)^2-(1-w_0/v_0)^2}
-(1+w_0/v_0)]/2$. In the limit of vanishingly small $\tau_\phi$, the
random orientation factor in Eq.\ (\ref{LE1}), $\cos \phi(t)$, can be
replaced by an effective noise source with zero mean and
autocorrelation function $\langle \cos \phi(t) \cos \phi(0) \rangle
\simeq 2\delta(t)/(2D_\phi)$ \cite{Ghosh}, so that the dynamics of
$x'=x-ut$ is well described by the effective 1D LE,
\begin{equation} \label{LEdiff}
\dot x' = -u+ v(x')\;\xi_\phi(t)/D_\phi\sqrt{2}
+\sqrt{D_0}\;\xi_x(t),
\end{equation}
where the multiplicative stochastic term on the r.h.s. must be
interpreted in Stratonovitch sense \cite{HT,Marchesoni}.

\subsection{The Fokker-Planck formalism}
For a more detailed analysis of the swimmer's stochastic dynamics we turn to the Fokker-Planck
equation (FPE) associated with the model Eqs.\ (\ref{LE1})
\cite{HT,Risken},
\begin{alignat}{2}
\label{FPE}
\partial_t P(\mathbf{r},\phi,t)&=-\nabla_iJ_i({\bf r},\phi,t),\\
\mathbf{J}(\mathbf{r},\phi,t)=&\left(\begin{array}{l}-D_0\partial_x+v(x)\cos\phi-u\\-D_0\partial_y+v(x)\sin\phi\\
-D_\phi\partial_\phi+\Omega\end{array}\right)P(\mathbf{r},\phi,t),\nonumber
\end{alignat}
and ${\bf\nabla}= (\partial_x, \partial_y, \partial_\phi)$. Here,
$\mathbf{r}=(x,y)$ denotes the particle's spatial coordinates in the
ADW moving frame, $x-ut\to x$ and $y \to y$ (the prime sign used in
Eq.\ (\ref{LE2wash}) has been dropped for simplicity). The 3D
functions $P$ and $\mathbf{J}$ denote, respectively, the probability
density and current of the particle in the state $(\mathbf{r},\phi)$
at time $t$. The FPE (\ref{FPE}) was numerically integrated by
combining the method of lines \cite{Schiesser} with a second-order
backward-difference scheme \cite{Hairer}. Periodic boundary
conditions were assumed for all three variables $x$, $y$ and $\phi$,
with relevant periods $L$, $L$ and $2\pi$. This amounts to the
one-zone reduced formulation of probability density \cite{Schmid}
\begin{equation}
\hat{P}(\mathbf{r},\phi,t)=\sum\limits_{k,n,m=-\infty}^\infty
P(x+kL,y+nL,\phi+2\pi m,t).
\end{equation}
The particle's drift velocities in the ADW frame are then computed as
the stationary currents in the $x$ and $y$ direction,
\begin{equation}
v_{x,y}=\int\limits_0^L\mathrm{d}x\int\limits_0^L\mathrm{d}y\int\limits_0^{2\pi}
\mathrm{d}\phi\,\lim\limits_{t\to\infty}J_{x,y}(\mathbf{r},\phi,t).
\label{vdexact}
\end{equation}
The final result for $v_x$ is obtained by transforming it back to the
laboratory frame,  $v_x\to v_x+u$. For achiral particles, $\Omega=0$,
the transverse coordinate $y$ is dispensable, so that, upon
integration over $y$, the FPE (\ref{FPE}) is immediately reduced to a
partial differential equation for the 2D probability density
$P(x,\phi,t)$. In the following we shortly address two limiting cases
of such a reduced FPE, corresponding, respectively, to the two-state
model and the reduced 1D model approximations of the LEs (\ref{LE1})
with $\Omega=0$, as introduced above.

{\it Ballistic approximation.} In the ballistic regime, the
particle's orientation is almost constant during its spatial
relaxation. Accordingly, the FPE can be further reduced to a 1D
partial differential equation in $x$ for a fixed, but arbitrary $\phi$. The stationary reduced probability current has a manageable
expression as a function of $\phi$ \cite{Schmid}, i.e.
\begin{alignat}{2}
\hat{J}&(\phi)=\frac{D_0}{L^2}\left\{1-\exp\left[-\frac{Lv_0}{D_0}\left(\frac{v_0+w_0}{2v_0}\cos\phi-\frac{u}{v_0}\right)\right]\right\}\nonumber\\
&\times\left\{\int\limits_0^1\mathrm{d}\xi\int\limits_0^1\mathrm{d}\zeta\,\exp\left[-\frac{Lv_0}{D_0}\left(\left(\frac{v_0+w_0}{2v_0}\cos\phi-\frac{u}{v_0}\right)\zeta\right.\right.\right.\nonumber\\
&\left.\left.\left.-\frac{v_0-w_0}{4\pi
v_0}\cos\phi\Big(\sin(2\pi(\zeta+\xi))-\sin(2\pi\xi)\Big)\right)\right]\vphantom{\int\limits_0^1}\right\}^{-1},
\label{Jball}
\end{alignat}
which, upon averaging over a uniform $\phi$ distribution with $\phi
\in [0,2\pi]$ and transforming back to the laboratory frame, yields
the most accurate estimates of the drift velocity in the ballistic
regime,
\begin{equation} \label{vxballistic}
v_x=\frac{L}{2\pi}\int\limits_0^{2\pi}\mathrm{d}\phi\,\hat{J}(\phi)+u.
\end{equation}

{\it Diffusive approximation.} In the opposite limit of fast angular
relaxation,  the dependence on the coordinate $\phi$ can be projected
out by means of a mapping procedure \cite{Kalinay} to be detailed in
an upcoming technical report. In leading order of the perturbation
parameter $l_\phi/L$, we  obtained a 1D partial differential equation
\begin{equation} \label{FPEdiff}
\partial_t P(x,t)=\partial_x^2 a(x)P(x,t)-\partial_x b(x)P(x,t),
\end{equation}
with $a(x)=v^2(x)/2D_\phi +D_0$ and $b(x)=(d/dx) v^2(x)/4D_\phi-u$.
This is the FPE corresponding to the reduced multiplicative process
derived in Eq.\ (\ref{LEdiff}). Upon computing the stationary
probability current,
\begin{alignat}{2}
\hat{J}=&\left[1-\exp\left(-\int\limits_0^L\mathrm{d}x\,\frac{b(x)}{a(x)}\right)\right]\nonumber\\
\times&\left[\int\limits_0^L\mathrm{d}x\int\limits_0^L\mathrm{d}y\,\frac{1}{a(x)}\exp\left(\int\limits_x^{y+x}\mathrm{d}z\,\frac{b(z)}{a(z)}\right)\right]^{-1},
\end{alignat}
\newpage
the drift velocity is finally expressed as $v_x=L\hat{J}+u$.

\section*{Acknowledgments}

This work has been supported by the cluster of excellence Nanosystems Initiative Munich (PH). PH and
FM acknowledge a financial support from the Center for Innovative
Technology (ACIT) of the University of Augsburg. SS thanks the
Alexander von Humboldt Stiftung for granting him a Bessel Research
Award.

\end{document}